\documentclass[a4paper, oneside, twocolumn, notitlepage, 10pt]{extarticle_ecoc}
\usepackage{ecoc}

\usepackage{amsmath,amssymb}
\usepackage{pgfplots}
\pgfplotsset{compat=1.18}
\usepgfplotslibrary{colormaps}
\usetikzlibrary{spy}
\usetikzlibrary{shapes.geometric}
\usepackage{subcaption}
\usepackage{graphicx}
\usepackage{caption}

\usepgfplotslibrary{external} 
\usepackage{comment}
\usepackage{siunitx}
\usepackage{wrapfig}
\usepackage{mathtools}
\usepackage[cal=cm]{mathalfa}

\definecolor{new_blue}{RGB}{93,169,233}
\definecolor{new_green}{HTML}{0FA3B1}
\definecolor{new_red}{HTML}{DD2D4A}
\definecolor{new_braun}{HTML}{79745C}
\definecolor{new_cherry}{HTML}{880D1E}
\definecolor{new_fildisi}{HTML}{F5EFED}
\definecolor{new_turqouise}{HTML}{508991}
\definecolor{new_purple}{HTML}{473198}
\definecolor{new_gray}{HTML}{67597A}

\definecolor{new_purple_soft}{HTML}{ACA2D2}
\definecolor{new_gray_soft}{HTML}{BCB5C4}
\definecolor{new_red_soft}{HTML}{DE828E}

\definecolor{KIT_blue}{RGB}{70,100,170}
\definecolor{KIT_green}{RGB}{0,150,130}
\definecolor{KIT_red}{RGB}{162,34,35}

\newcommand{\FER}{$\mathrm{FER}$ }
\newcommand{\SKR}{$\mathrm{SKR}$ }
\newcommand{\VA}{$V_{\mathrm{A}}$ }

\begin{document}
\selectlanguage{english}    %

\title{Parameter Optimization of Rate-Adaptive Continuous-Variable Quantum Key Distribution Systems}%

\author{
    Erdem Eray Cil\textsuperscript{(1)}, Jonas Berl\textsuperscript{(1,2)}, Laurent Schmalen\textsuperscript{(1)}
}

\maketitle                  %

\begin{strip}
    \begin{author_descr}

        \textsuperscript{(1)} Communications Engineering Lab (CEL), Karlsruhe Institute of Technology (KIT),
        \textcolor{blue}{\uline{erdem.cil@kit.edu}}

        \textsuperscript{(2)} Adva Network Security GmbH%
    \end{author_descr}
\end{strip}

\renewcommand\footnotemark{}
\renewcommand\footnoterule{}

\begin{strip}
    \begin{ecoc_abstract}
        We propose an optimization method for rate-adaptive CV-QKD systems, improving the \SKR by up to $15\%$. A single information reconciliation setup can generate secret keys up to a distance of $\qty{112}{km}$. This enables a unified reconciliation system, thereby facilitating the commercialization of CV-QKD. \textcopyright 2024 The Author(s)
    \end{ecoc_abstract}
\end{strip}

\section{Introduction}

Quantum key distribution (QKD) has emerged as a physical layer security solution, providing secure keys for symmetric encryption schemes. Among QKD techniques, continuous-variable quantum key distribution (CV-QKD) has demonstrated significant potential for secure communication over long distances, as evidenced by the successful exchange of cryptographic keys over distances exceeding $\qty{200}{km}$~\cite{208kmCV-QKD}.

In CV-QKD systems, the measured quantum states deviate from the transmitted states due to channel noise, eavesdropping, or imperfections in the transmitter/receiver setup. To establish a common key, an information reconciliation (IR) step is required, which employs forward error correction (FEC) mechanisms to mitigate the impact of the noise and the above-mentioned deviations. To ensure operability across various fiber lengths or unstable channels, the IR step must adapt the rate of the error-correcting code for optimal performance. Several methods have been proposed for rate adaptation, including raptor codes~\cite{RaptorCodes} and raptor-like \mbox{low-density parity-check}~(RL-LDPC) codes~\cite{RL_Code1, RL_Code2}, as well as punctured or shortened fixed-rate codes~\cite{Puncturing_Shortening}. Recently, we introduced a protograph-based RL-LDPC code construction method that enables the generation of codes with robust performance across a wide range of rates~\cite{Cil_Rate-Adaptive_Protograph-Based_Raptor-Like, SPPCOM}.

To achieve optimal performance in a CV-QKD system, adjustable parameters must be fine-tuned. The parameter optimization for fixed-rate systems has been studied in~\cite{curve_fitting}. To ensure the system operates within the same signal-to-noise ratio ($\mathrm{SNR}$) region as the fixed-rate code, modulation variance is adjusted. After determining the IR performance of the fixed-rate code, the modulation variance is optimized to maximize the secret key rate ($\mathrm{SKR}$).

However, optimizing the $\mathrm{SKR}$ for a system with rate-adaptive IR poses a challenge, as the rate-adaptive code operates over a wide range of $\mathrm{SNR}$ values. Moreover, for each value of $\mathrm{SNR}$, there is a wide range of possible code rate values that can be used in IR. This introduces an additional dimension to the $\mathrm{SKR}$ optimization problem, necessitating the optimization of both the $\mathrm{SNR}$ and the code rate $R$. To the best of our knowledge, joint optimization of these parameters for rate-adaptive IR has not been previously investigated in the literature.

In this paper, we propose a parameter optimization method to maximize the $\mathrm{SKR}$ in CV-QKD. By employing the proposed optimization method, distances up to $\qty{112}{km}$ become feasible with a single FEC code, significantly simplifying IR for distance-adaptive CV-QKD systems.

\begin{figure*}[t] 
\begin{minipage}{\dimexpr\textwidth-2\fboxsep-2\fboxrule\relax}
    \begin{subfigure}{0.5\textwidth}
      \centering
\includegraphics{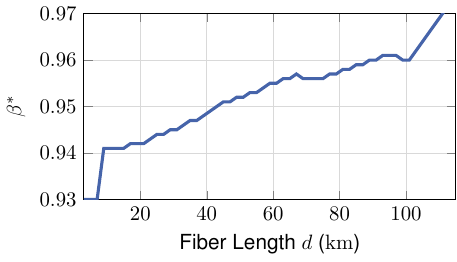}
 \centering 
  \captionsetup{margin={4.5cm,0cm}, justification=raggedright, singlelinecheck=false}
  \caption{}
 \label{fig:sub_beta}
 \end{subfigure}
 \begin{subfigure}{0.5\textwidth}
       \centering
       \includegraphics{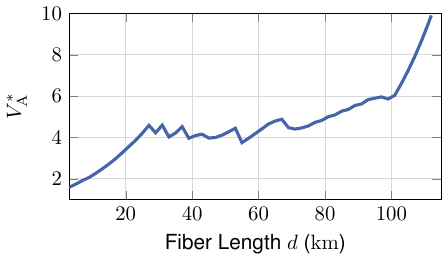}
   \captionsetup{margin={4.5cm,0cm}, justification=raggedright, singlelinecheck=false}
  \caption{}
 \label{fig:sub_VA}
 \end{subfigure}
 
 \vspace{0.25cm}
 
   \begin{subfigure}{0.5\textwidth}
     \centering
     \includegraphics{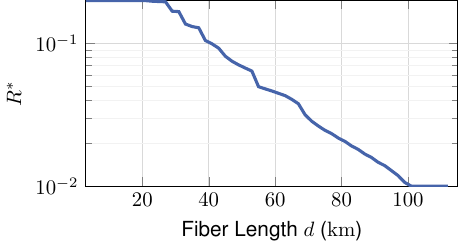}
  \captionsetup{margin={4.5cm,0cm}, justification=raggedright, singlelinecheck=false}
  \caption{}
 \label{fig:sub_rate}
 \end{subfigure}
 \begin{subfigure}{0.5\textwidth}
        \centering
        \includegraphics{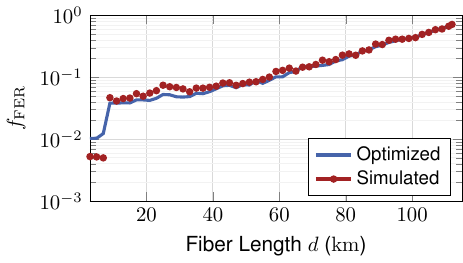}
 \captionsetup{margin={4.5cm,0cm}, justification=raggedright, singlelinecheck=false}
  \caption{}
 \label{fig:sub_FER}
 \end{subfigure}
 \end{minipage}
 \vspace{0.4cm}

 \caption{Jointly optimized values of reconciliation efficiency ($\beta$), modulation variance ($V_\mathrm{A}$), the code rate ($R$) and resulting frame error rate~($f_\mathrm{FER}$) for the fiber lengths~($d$) ranging from $\qty{3}{km}$ to $\qty{112}{km}$.}
 
 \label{fig:opt_results_3_figure}
\end{figure*}

\section{Parameter Optimization}

Typically, in the design of CV-QKD systems, the primary objective is to maximize the number of secret keys generated. To accomplish this, we aim to select the parameters that maximize the $\mathrm{SKR}$. By assuming reverse reconciliation and ignoring finite-size effects, the $\mathrm{SKR}$ reads~\cite{Milicevic2018}
\begin{equation}
\label{eq:SKR} 
\mathrm{SKR}= (1 - \mathrm{FER}) (\beta {I_\mathrm{AB}} - {\chi_\mathrm{BE}}).
\end{equation}

Here, $\mathrm{FER}$ represents the frame error rate after decoding, $I_\mathrm{AB} \geq 0$ denotes the capacity of the quantum channel, and $\chi_\mathrm{BE} \geq 0$ represents the Holevo bound on the information leaked to the eavesdropper.

To express the $\mathrm{SKR}$ as a function of the adjustable parameters in the system, namely the modulation variance~$V_\mathrm{A}$ and the reconciliation efficiency~$\beta$, we fix the other system parameters. In this case, $I_\mathrm{AB}$ and $\chi_\mathrm{BE}$ are monotonically increasing functions of $V_\mathrm{A}$. The $\mathrm{FER}$ depends on both $\beta$ and $V_\mathrm{A}$. For a fixed $V_\mathrm{A}$, the $\mathrm{FER}$ is a monotonically non-decreasing function of $\beta$, whereas it is monotonically non-increasing with $V_\mathrm{A}$ for a fixed $\beta$. Due to this relationship, both the values $V_\mathrm{A}$ and $\beta$ must be optimized jointly.

To determine the optimal parameters $(V_\mathrm{A}^*,\beta^*)$ that maximize the \SKR given by (\ref{eq:SKR}), we require the dependencies of $I_\mathrm{AB}$, ${\chi_\mathrm{BE}}$ and $\mathrm{FER}$ on \VA and $\beta$. We assume heterodyne transmission with Gaussian states and a trusted receiver scenario, which allows us to analytically calculate the ${\chi_\mathrm{BE}}$ and $I_\mathrm{AB}$ as in~\cite{Laudenbach}. To obtain the \FER characteristics, we run Monte Carlo (MC) simulations for the protograph-based RL-LDPC code~\cite{SPPCOM, Cil_Rate-Adaptive_Protograph-Based_Raptor-Like} using the sum-product decoding algorithm. We assume that multidimensional reconciliation~\cite{leverrierMultidimensionalReconciliationContinuousvariable2008} transforms the channel seen by the FEC to a binary-input additive white Gaussian noise (BI-AWGN) channel. Note that FEC performance depends on the $\mathrm{SNR}$ $s$ of the received signal and the code rate $R$. Thus, we directly set $s$ and $R$ parameters in the MC simulations. By converting $(V_A, \beta)$ to $(s,R)$, we can utilize the same simulation results for all parameter settings.

\begin{figure*}[b]
    \noindent\rule{\linewidth}{0.4pt}

    \begin{equation}
        (V_\mathrm{A}^*, \beta^*) = \mathrm{arg} \ \underset{(V_\mathrm{A}, \beta)}{\mathrm{max}} \left(1 - f_\mathrm{FER}\left( f_\mathrm{SNR}(V_\mathrm{A}), \frac{\beta}{2} f_\mathrm{I_\mathrm{AB}}\left(V_{\mathrm{A}}\right) \right) \right)\cdot \big( {\beta} f_\mathrm{I_\mathrm{AB}}\left(V_{\mathrm{A}}\right) \vspace{-2pt} - \vspace{-2pt} f_\mathrm{\chi_{BE}}\left(V_\mathrm{A}\right)\big) \label{eqn:optimization}
    \end{equation}
\end{figure*}

As simulating the rate-adaptive code for each $(s, R)$ pair is not feasible, we select discrete rate values $\underline{R} = \{R_1, R_2, \dots \}$ for the MC simulations. Afterwards for each rate $R_i$, we fit a polynomial to each simulation result $f_\mathrm{FER}(s,R_i)$, following the approach in~\cite{curve_fitting}. Subsequently, we interpolate the polynomial coefficients obtained for the discrete rate values to obtain a closed-form expression $f_\mathrm{FER}(s,R)$ for $\mathrm{FER}$.

After obtaining the \FER characteristics, we have a closed-form equation for the \SKR that can be utilized in the optimization problem stated in (\ref{eqn:optimization}). The functions $f_{\chi_\mathrm{BE}}(V_\mathrm{A})$ and $f_\mathrm{I_\mathrm{AB}}(V_{\mathrm{A}})$ represent the dependency of the Holevo bound and the channel capacity on $V_\mathrm{A}$ \cite{Laudenbach}. Note that $\beta$ is defined as in~\cite{beta_defn}, as the reconciliation process is carried out after separating the two quadratures following heterodyne detection. We can solve the optimization problem in (\ref{eqn:optimization}) either using non-linear optimization tools or by sweeping the parameters $(V_\mathrm{A},\beta)$. The latter is feasible due to the closed-form expression obtained for the objective function.

\section{Simulation Results}

\begin{figure*}[t]
    \centering
    \includegraphics{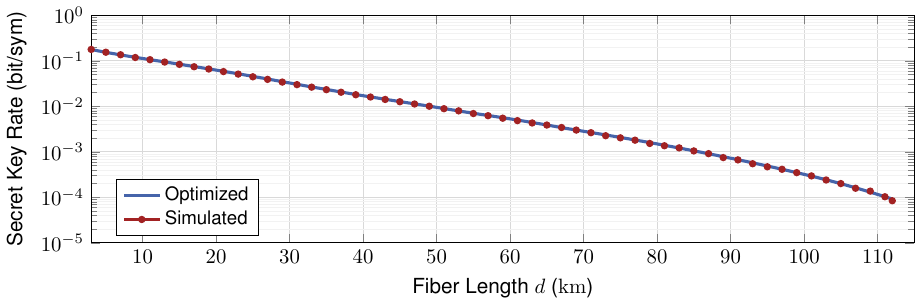}
\caption{ Optimized secret key rate ($\mathrm{SKR}$) values as a function of the fiber length ($d$), along with simulated \SKR values to validate the proposed optimization approach.}
    \label{fig:SKR}
\end{figure*}

To evaluate the efficiency of the proposed optimization method for rate-adaptive reconciliation, we optimize the parameters $(V_\mathrm{A},\beta)$ for a CV-QKD system and subsequently simulate the IR step. In the simulation, we model the synthetic post-multidimensional reconciliation channel as a BI-AWGN channel. We consider a quantum efficiency of the detector of $\eta=0.55$, a fiber attenuation of $\alpha=\qty{0.2}{dB/km}$, and practical noise values derived from experimental data in \cite{experimental_paper}: an excess noise level at the channel input of $\xi_{\mathrm{ch, A}}=0.05 $ shot noise units~($\mathrm{SNU}$) and electronic noise in the detector of \mbox{$\xi_\mathrm{rec}=\qty{0.18}{SNU}$}. We optimize the parameters $(V_\mathrm{A}, \beta)$ for fiber lengths ranging from $\qty{3}{km}$ to $\qty{112}{km}$ for $V_\mathrm{A}\in [0.5,10]$.

Figure \ref{fig:opt_results_3_figure} presents the optimized parameters $(V_\mathrm{A}^*, \beta^*)$ and the corresponding \FER values. The protograph-based RL-LDPC code is optimized for rates ranging from $0.01$ to $0.2$. As illustrated in Fig.~\ref{fig:opt_results_3_figure}(b), when the system operates within this range, the value of \VA remains between $4$ and $6$. This suggests that the optimization algorithm prefers a stable \VA value and increases $\beta$ with increasing distance, although higher $\beta$ values result in degraded FEC performance, as observed in Fig.~\ref{fig:opt_results_3_figure}(d). The increase in $\beta$ for a fixed value of \VA can be attributed to the decrease in the multiplicative gap between $I_\mathrm{AB}$ and $\chi_\mathrm{EB}$ as the distance increases.

To verify the accuracy of our approach to represent the \FER in a closed-form expression, we simulate the protograph-based RL-LDPC code using the optimized parameter values. In Fig.~\ref{fig:opt_results_3_figure}(d), the \FER values obtained from optimization are plotted alongside the \FER values from the simulation with the optimized parameters. The simulated IR with optimized parameter values closely match the\FER values, validating the optimization.

The systems operating at $R=0.2$ and $R=0.01$ provide insight into the operation of IR at two extreme points. Higher rates are preferred for shorter distances, with lower \FER values around $0.04$. In contrast, longer distances require lower rates with \FER values significantly higher than $0.1$. This observation may enable the development of decoding algorithms or code designs specifically targeting CV-QKD IR, as the algorithms used today are optimized for classical communication scenarios with much lower target error rates. 

In Fig.~\ref{fig:SKR}, we show the predicted \SKR values from the optimization alongside the \SKR values obtained from the simulation. The values from the simulation correspond to the maximized values in the optimization, indicating that the optimization algorithm works as intended. Notably, the joint optimization of parameters to maximize the \SKR also extends the distance range within which the reconciliation algorithm operates reliably. Figure~\ref{fig:SKR} demonstrates that a single encoder implementation can effectively support distances up to $\qty{112}{km}$.

\begin{figure}[t!]
\includegraphics{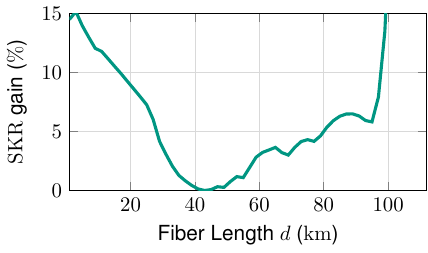}
    \caption{Improvement of the secret key rate ($\mathrm{SKR}$) achieved by proposed optimization compared to modulation variance ($V_\mathrm{A}$) optimization with fixed reconciliation efficiency $\beta=0.95$ for fiber length $d$ }
    \label{fig:improvement}
\end{figure}

Lastly, we compare our optimization approach to existing methods. For this purpose, we set the value of $\beta$ to $0.95$ as in~\cite{experimental_paper} and optimize the values of \VA for fiber lengths ranging from $\qty{3}{km}$ to $\qty{112}{km}$. We then compare the resulting \SKR values with the \SKR values obtained from the joint optimization. Figure~\ref{fig:improvement} shows the improvement achieved by our approach over the optimization of \VA alone. Our joint parameter optimization yields up to a $15\%$ increase in $\mathrm{SKR}$, depending on the distance $d$. For distances around $\qty{40}{km}$, the optimized $\beta$ value is approximately $0.95$, which explains why there is no observed gain over choosing $\beta=0.95$. However, for distances exceeding $\qty{100}{km}$, the \SKR for $\beta=0.95$ approaches $0$, resulting in an infinite \SKR improvement.

\section{Conclusion}
In this paper, we present a parameter optimization method for rate-adaptive CV-QKD systems. This approach improves the \SKR up to $15\%$  compared to optimization of the modulation variance alone. By integrating this method with the previously designed protograph-based RL-LDPC code, secret keys can be generated for distances up to $\qty{112}{km}$ using a single IR setup. This enables the development of a unified IR system capable of operating efficiently across a wide range of distances, thereby facilitating the commercialization of CV-QKD systems.

\section{Acknowledgements}
This work was funded by the German Federal Ministry of Education and Research (BMBF) under grant agreements 16KISQ056 and 16KISQ052K~(DE-QOR).

\printbibliography%

\vspace{-4mm}

\end{document}